\documentclass[fleqn,usenatbib]{mnras}

\usepackage{newtxtext,newtxmath}

\usepackage[T1]{fontenc}

\DeclareRobustCommand{\VAN}[3]{#2}
\let\VANthebibliography\thebibliography
\def\thebibliography{\DeclareRobustCommand{\VAN}[3]{##3}\VANthebibliography}

\usepackage{slashbox}
\usepackage{graphicx}	
\usepackage{amsmath}	
\usepackage[dvipsnames]{xcolor}
\usepackage{multirow}
\usepackage{array}
\usepackage{comment}

\usepackage{pgfplots}
\usepackage{comment}

    \usetikzlibrary{intersections}
    \pgfplotsset{compat=1.11}
\pgfplotsset{
    every axis plot/.append style = {font = \normalsize}
  }
  
\newcolumntype{x}[1]{wc{#1}}
\newcolumntype{M}[1]{>{\centering\arraybackslash}m{#1}}

\title[Temporal polarization of accretion flow]{Temporal polarization properties of accretion flows 
into supermassive black holes: INSANE as an alternative}

\author[Long et al.]{
Jane SiNan Long$^{1,2}$\thanks{E-mail: sinan.long.23@ucl.ac.uk (JSL), akhil\_uniyal@sjtu.edu.cn (AU)}, 
Akhil Uniyal$^{2,3}$, 
Yosuke Mizuno$^{2,4,5,6}$, 
Koushik Chatterjee$^{7}$  
and Kinwah Wu$^{1,8,9,10}$
\\
$^{1}$Mullard Space Science Laboratory, 
  University College London, 
  Holmbury St Mary, Surrey, RH5 6NT, United Kingdom \\
$^{2}$Tsung-Dao Lee Institute \& 
  School of Physics and Astronomy, 
  Shanghai Jiao Tong University, Shanghai 201210, China \\
$^{3}$Institute of Fundamental Physics and Quantum Technology, \& School of Physical Science and Technology, Ningbo University, Ningbo, Zhejiang 315211, China\\
$^{4}$School of Physics and Astronomy, Shanghai Jiao Tong University, Shanghai 200240, China \\ 
$^{5}$Key Laboratory 
  for Particle Astrophysics and Cosmology (MOE) 
  and Shanghai Key Laboratory for Particle Physics 
  and Cosmology, \\ 
  \hspace*{0.5em}Shanghai Jiao Tong University, Shanghai 200240, China  \\ 
  $^{6}$Institut f\"ur Theoretische Physik, Goethe-Universit\"at Frankfurt, D-60438 Frankfurt am Main, Germany \\
  $^{7}$Department of Physics, University of Maryland, 7901 Regents Drive, College Park, MD 20742, USA \\
$^{8}$Department of Physics, Chinese University of Hong Kong, 
 Shatin, NT, Hong Kong SAR, China \\
$^{9}$Kavli Institute for the Physics and Mathematics of the Universe (WPI), UTIAS, The University of Tokyo, Kashiwa, Chiba 277-8583, Japan   \\ 
$^{10}$Research School of Astronomy and Astrophysics, Australian National University, 
 Canberra, ACT2611, Australia
}

\date{Accepted XXX. Received YYY; in original form ZZZ}

\pubyear{\the\year{}}

\begin{document}
\label{firstpage}
\pagerange{\pageref{firstpage}--\pageref{lastpage}}
\maketitle

\begin{abstract}

  Temporal variability and polarization provide unique diagnostics of the magnetic field structure and plasma dynamics in accretion flows around supermassive black holes. We investigate the temporal polarization properties of horizon-scale emission using general relativistic magnetohydrodynamic (GRMHD) simulations coupled with polarized general relativistic radiative transfer (GRRT) for three accretion states: Standard and Normal Evolution (SANE), Magnetically Arrested Disk (MAD), and an intermediate regime (INSANE). 
 Fourier polarization-mode decomposition, through the ratio $|\beta_m|/|\beta_2|$, shows that MAD produces the most rotationally symmetric EVPA patterns, SANE the most asymmetric, and INSANE lies in between. 
The multi-scale temporal behaviour of the intensity is investigated using the Continuous Wavelet Transform and Hilbert--Huang Transform. Both methods identify localized oscillatory features associated with magnetic-flux eruption events in the MAD simulations. 
In the shorter interval examined, INSANE distributes its variability more evenly among several time-scales, while the SANE behaviour depends more strongly on the electron-temperature prescription. To quantify the spectral distribution, we define a fast power fraction $F_{\mathrm{fast},I}$ for periods of $80$--$200\,M$ within a broad band of $80$--$1500\,M$. Both methods give the smallest fractions for MAD; the ordering of state medians, INSANE $>$ SANE $>$ MAD, persists when the upper fast-band boundary is varied from 200 to $500\,M$.
Our analysis demonstrates that combining polarization morphology with multi-scale temporal analysis provides a useful means of distinguishing magnetization states, supporting INSANE as a distinct observational alternative to conventional MAD/SANE for interpreting horizon-scale observations of M87$^\ast$ and related systems.

\end{abstract}




\section{Introduction}

Black holes accreting at low Eddington ratios power some of the most energetic relativistic jets in the Universe, yet the physical state of the plasma in the immediate vicinity of the event horizon remains uncertain \citep{Yuan:2014gma}. A key unresolved question concerns the magnetic flux threading the black hole and inner accretion flow properties, which determine whether the system operates in a Standard And Normal Evolution (SANE) configuration, a Magnetically Arrested Disk (MAD) state, or an intermediate regime with partial magnetic flux accumulation \citep{narayan2012grmhd}. This distinction has profound implications for jet launching, particle acceleration, and the temporal variability of the emitted radiation.

The nearby galaxy M87 provides an exceptional laboratory for studying these processes. With horizon-scale imaging achieved by the Event Horizon Telescope (EHT) \citep{EHTM87First:2019dse, EHTSgrAFirst:2022wkp}, combined with multi-wavelength monitoring and polarization measurements, M87 offers direct observational access to plasma dynamics occurring on horizon-scale variability \citep{EventHorizonTelescope:2021bee, EventHorizonTelescope:2021srq}. 
A complementary perspective is provided by observations of Sgr~A*, the supermassive black hole at the centre of the Milky Way, where rapid intrinsic variability and lower jet power suggest a different accretion environment and potentially a lower net magnetic flux state \citep{EventHorizonTelescope:2022exc, EventHorizonTelescope:2022urf}. While several EHT modelling efforts for M87 indicate that MAD configurations reproduce many observed features, particularly jet energetics and polarized structure, current observational constraints do not uniquely exclude SANE or intermediate-magnetization (“weak-MAD”) flows \citep{EventHorizonTelescope:2021bee, EventHorizonTelescope:2021srq, EventHorizonTelescope:2025dua, EventHorizonTelescope:2025vum}. In contrast, for Sgr~A*, present observational and modelling studies generally favour SANE or moderately magnetized configurations, with strongly arrested MAD states appearing less consistent with the observed variability and polarization properties \citep{EventHorizonTelescope:2022exc, EventHorizonTelescope:2022urf, EventHorizonTelescope:2024hpu}. Together, these systems highlight the importance of variability diagnostics for distinguishing among different magnetic flux regimes in low-luminosity accreting black holes.

General Relativistic Magnetohydrodynamic (GRMHD) simulations predict distinct variability signatures associated with different magnetization states. SANE disks typically exhibit turbulence-dominated, small-scale stochastic fluctuations, whereas MAD systems display more coherent large-scale magnetic structures that can modulate emission on longer and often correlated timescales \citep{Narayan:2003by, Igumenshchev:2003rt, Igumenshchev:2007bh, Begelman:2021ufo}. Intermediate or partially arrested disks, sometimes referred to as INSANE or semi-MAD states, may produce variability characteristics that lie between these two extremes, complicating the interpretation of observational data \citep{Raha:2025bef}. Consequently, systematic comparisons between variability predictions of SANE, intermediate, and MAD simulations are essential for constraining the magnetic state of the M87 accretion flow.

Accretion flows forming the horizon-scale environments of black holes are intrinsically time-dependent, driven by magnetohydrodynamic turbulence, magnetic reconnection, and dynamical plasma instabilities. At millimeter wavelengths, particularly near 1.3 mm, the observed emission is dominated by synchrotron radiation produced by relativistic electrons energized within these turbulent magnetized flows \citep{EHTC2019_Interpretation}. Because the emitting plasma evolves on orbital and inflow timescales close to the innermost stable circular orbit (ISCO), the brightness distribution of the source is expected to vary on characteristic dynamical timescales that depend strongly on the black hole mass and spin \citep{Bardeen:1972fi}.

For the Galactic center black hole Sgr~A*, these characteristic timescales are on the order of minutes to tens of minutes, implying that appreciable structural variability can occur even during a single observing night, complicating the interpretation of time-averaged images. In contrast, the substantially larger mass of the M87 black hole shifts the relevant dynamical timescales to days or weeks, suggesting that short-duration observing campaigns should capture comparatively modest structural evolution, while longer baselines may reveal measurable changes in the morphology and azimuthal brightness distribution of the emission ring \citep{Wielgus2020}. Indeed, multi-epoch very-long-baseline interferometry (VLBI) observations spanning several years have already indicated the presence of long-term structural variability in M87, consistent with expectations from turbulent accretion dynamics.

The 2017 EHT observing campaign of M87, which covered multiple nights separated by several days, sampled only a few characteristic dynamical times near the ISCO. Consequently, only moderate day-to-day variations in ring thickness and brightness asymmetry were inferred, highlighting the importance of time-dependent modeling for interpreting interferometric observables. Because the EHT measures complex visibilities corresponding to the Fourier components of the sky brightness distribution, temporal evolution in the accretion flow is encoded directly in the variability of visibility amplitudes, phases, and polarization quantities rather than solely in reconstructed images \citep{EHTC2019_Variability}.

In this work, we employ time-dependent general relativistic magnetohydrodynamic (GRMHD) simulations spanning SANE, intermediate (INSANE), and MAD accretion states, coupled with polarized general relativistic radiative transfer (GRRT) calculations tailored to the physical parameters of M87. Using these simulations, we investigate how short- and intermediate-timescale variability in both total intensity and polarized emission manifests in interferometric observables and assess the extent to which variability diagnostics can distinguish between different magnetization regimes of the M87 accretion flow.

\section{Numerical setup}
\label{sec:method}

\subsection{General relativistic magnetohydrodynamic setup}

We conducted a set of three-dimensional (3D) GRMHD simulations of a magnetized torus around a spinning black hole using the {\tt H-AMR} code \citep{H-AMR}. Each simulation is initialized from a Fishbone–Moncrief hydrodynamic equilibrium torus \citep{Fishbone1976ApJ, Uniyal:2024sdv} with parameters $r_{\rm in} = 20 r_{\rm g}$ and pressure maximum location $r_{\rm max} = 41 r_{\rm g}$, where $r_{\rm g} \equiv GM/c^2 $ is the gravitational radius of the black hole and $M$ its mass for the MAD, INSANE and SANE cases while additional runs have also been performed with $r_{\rm in} = 6 r_{\rm g}$ and $r_{\rm max} = 12 r_{\rm g}$ for the SANE and INSANE cases which are considered standard choices in the literature. We use an ideal-gas equation of state with the adiabatic index $\Gamma_{\rm g} =4/3$ \citep{Rezzolla2013rehy}. Furthermore, the simulations employ a fixed mesh refinement strategy corresponding to an effective resolution of $384 \times 192 \times 192$. We summarize all GRMHD and radiative transfer models in Table~\ref{table:sum}. The details of the initial configuration setup can be found in Appendix~\ref{app: GRMHD}.

\begin{table*}
\centering
\begin{tabular}{cccccccccccc}
\hline
$a_*$ & ID & $r_{\rm in}$ & $r_{\rm max}$ & Rlow & Rhigh & $i$ & Munit$/10^{25}$  & $t_s$ & $t_f$ & $\Delta t$ & Resolution  \\
 &  & ($R_g$) & ($R_g$) &  & &   (deg) &    ($g$) & (M) & (M) & (M) &    \\
\hline
0.9375 & MAD\_20 & 20 & 41 &  1 & (10, 80) & 163 & (1.01, 2.62)  & 15000 & 30000 & 10 & $384 \times 192 \times 192$   \\
0.9375 & INSANE\_20  & 20 & 41 &  1 & (10, 80) & 163 & (9.42, 25.1)  & 15000 & 30000 & 10 & $384 \times 192 \times 192$   \\
0.9375 & INSANE\_6  & 6 & 12 &  1 & (10, 80) & 163 &  ($1.94 \times 10^{3}$, $5.41 \times 10^{3}$)   & 15000 & 30000 & 10 & $384 \times 192 \times 192$   \\
0.9375 & SANE\_20 & 20 & 41 &  1 & (10, 80) & 163 & (30.0, 86.1)  & 15000 & 30000 & 10 & $384 \times 192 \times 192$   \\
0.9375 & SANE\_6 & 6 & 12 &  1 & (10, 80) & 163 & ($2.47 \times 10^{3}$, $1.17 \times 10^{4}$)   & 15000 & 30000 & 10 & $384 \times 192 \times 192$   \\
0.5 & MAD\_0.5 & 20 & 41 &  1 & (10, 80) & 163 & (1.29, 3.27)  & 15000 & 30000 & 10 & $384 \times 192 \times 192$   \\
0.5 & INSANE\_0.5  & 20 & 41 &  1 & (10, 80) & 163 & (17.0, 41.9)  & 15000 & 30000 & 10 & $384 \times 192 \times 192$   \\
0.5 & SANE\_0.5 & 20 & 41 &  1 & (10, 80) & 163 & (85.6, 190.8)  & 15000 & 30000 & 10 & $384 \times 192 \times 192$   \\
\hline
\end{tabular}
\caption{Overview of GRMHD and radiative transfer models employed to produce various time series of synthetic images. \label{table:sum}}
\end{table*}

%
%

\begin{figure*}
\centering
\includegraphics[width=2.\columnwidth]{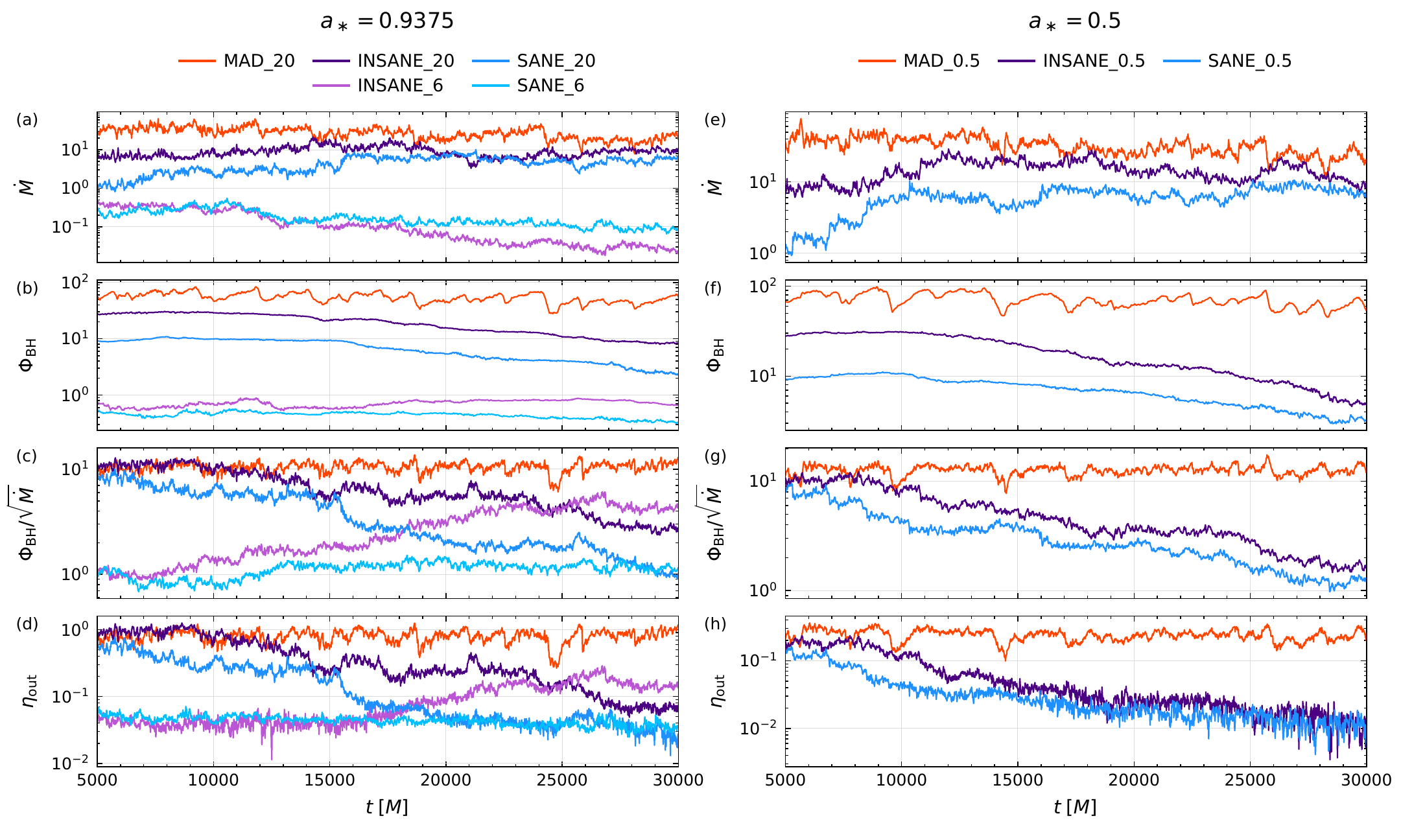}
\caption{Time evolution of the horizon mass-accretion rate
$\dot M$, horizon magnetic flux
$\phi$, normalized horizon magnetic flux
$\phi_{\rm BH}=\Phi_{\rm BH}/\sqrt{\dot M}$, and total outflow
efficiency $\eta_{\rm out}$. The left column shows models with
$a_*=0.9375$ and the right column models with $a_*=0.5$. Model
suffixes 20 and 6 denote $(r_{\rm in},r_{\rm max})=(20,41)\,R_g$
and $(6,12)\,R_g$, respectively. Only the large-torus configuration
is available for $a_*=0.5$. The curves illustrate that the ordering
of the dynamical quantities depends on both accretion state and
initial torus configuration.}
\label{fig:phi}
\end{figure*}

Figure~\ref{fig:phi} shows the horizon mass-accretion rate $\dot M$, the horizon magnetic flux $\Phi_{\rm BH}$, the normalized magnetic flux $\phi_{\rm BH}\equiv\Phi_{\rm BH}/\sqrt{\dot M}$, and the total outflow efficiency $\eta_{\rm out}$. At fixed large-torus geometry, MAD generally has the highest characteristic horizon flux and outflow efficiency, with INSANE typically between MAD and SANE, although individual curves can overlap. MAD remains in a saturated state throughout the evolution, with $\phi_{\rm BH}$ and $\eta_{\rm out}$ fluctuating around approximately constant values. Its episodic decreases of $\Phi_{\rm BH}$ and subsequent recovery are consistent with magnetic-flux eruptions followed by renewed inward advection.

The SANE and INSANE models show opposite trends for the two torus sizes. In the small-torus models (SANE\_6, INSANE\_6), $\dot M$ decreases with time while $\phi_{\rm BH}$ increases; in the large-torus models (SANE\_20, INSANE\_20), $\dot M$ increases or remains approximately constant while $\phi_{\rm BH}$ decreases. The absolute horizon flux does not follow these trends. In the small-torus models, $\Phi_{\rm BH}$ remains approximately constant (INSANE\_6) or decreases only slowly (SANE\_6), so the increase of $\phi_{\rm BH}$, by a factor of $\sim4$--$5$ in INSANE\_6, results almost entirely from the decrease of $\dot M$ and cannot be interpreted as a growth of the absolute horizon flux. In the large-torus models, $\Phi_{\rm BH}$ decreases by a factor of $\sim3$--$4$ after $t\approx1.3$--$1.5\times10^4\,M$ while $\dot M$ is maintained, so the decline of $\phi_{\rm BH}$ reflects a genuine loss of horizon flux, with an additional contribution from the rising $\dot M$ in SANE\_20.

These differences arise because mass and magnetic flux are accreted differently. The accretion rate traces the instantaneous supply of gas, whereas $\Phi_{\rm BH}$ is an accumulated quantity that changes only when magnetic flux of either polarity is advected inward or is removed by reconnection or escape. The small tori have their inner edge and pressure maximum close to the black hole and hence a small mass reservoir with a short inflow time; the torus is progressively depleted and $\dot M$ decreases, while the flux accumulated on the horizon at early times is largely retained. The large tori have a more extended reservoir and a longer inflow time, which sustain the accretion rate. Since the initial magnetic field follows the torus density distribution, this sustained inflow eventually delivers the outer part of the initial magnetic loop, which carries flux of opposite polarity; we attribute the decrease of $\Phi_{\rm BH}$ in these models to the partial cancellation of the previously accumulated horizon flux by this material \citep[see also][]{Beckwith:2007sr}.

The outflow efficiency follows the evolution of $\phi_{\rm BH}$, as expected for magnetically driven outflows. It is generally highest in MAD, increases from $\sim0.04$ to $\sim0.15$--$0.2$ in INSANE\_6, and decreases by about an order of magnitude in SANE\_20 and INSANE\_20, such that INSANE\_6 exceeds INSANE\_20 in both $\phi_{\rm BH}$ and $\eta_{\rm out}$ at $t\gtrsim2.4\times10^4\,M$. At late times, $\eta_{\rm out}$ is higher in INSANE\_6 than in SANE\_6. The high-spin MAD model occasionally reaches $\eta_{\rm out}>1$, indicating net extraction of black-hole rotational energy \citep{Tchekhovskoy2011MNRAS}.

\subsection{Polarised general relativistic radiative transfer setup}

In this work, we carry out GRRT calculations for the 3D accretion models using the \textsc{RAPTOR} code  \citep{RAPTOR1, RAPTOR2}. The simulated source is assumed to represent M87$^{\ast}$, with a black hole mass of $M = 6.5 \times 10^9 M_{\odot}$ and a distance of 16.8 Mpc \citep{EventHorizonTelescope:2019ggy}. Images are generated on a grid with a pixel resolution of 500 $\times$ 500, while the total field of view in the image plane is set to $130 \mu as$ ($=34 \ R_{\rm g}$). As commonly adopted in modeling horizon-scale emission from supermassive black holes, we consider synchrotron radiation from a purely thermal electron population in all simulations. The electron temperature is determined using the standard $R$--$\beta$ prescription for the ion-to-electron temperature ratio \citep{MoscibrodzkaApJ},
\begin{equation}
\frac{T_i}{T_e}=\frac{R_{\rm low}+R_{\rm high} \beta^{2}}{1+\beta^{2}},
\end{equation}
where $\beta = p_{\rm g}/p_{\rm m}$ denotes the plasma beta, defined as the ratio of gas pressure to magnetic pressure. We fix $R_{\rm low}=1$ and use $R_{\rm high}=10$ and $80$, giving two electron-temperature prescriptions for each GRMHD simulation. The viewing inclination is fixed at $i = 163^{\circ}$.

We considered the data between $t=15000$~M and $30000$~M, with a cadence of 10~M, unless otherwise specified. The density normalization is specified by the mass unit $M_{\rm unit}$ in Table~\ref{table:sum}. The analysed light curves are normalized to a median 230\,GHz flux density of $0.5$\,Jy, consistent with the intensity observed for M87$^{\ast}$. 

\section{Asymmetry of polarization morphology}
\label{sec:pol_morphology}

\begin{figure*}
\centering
\includegraphics[width=2.0\columnwidth]{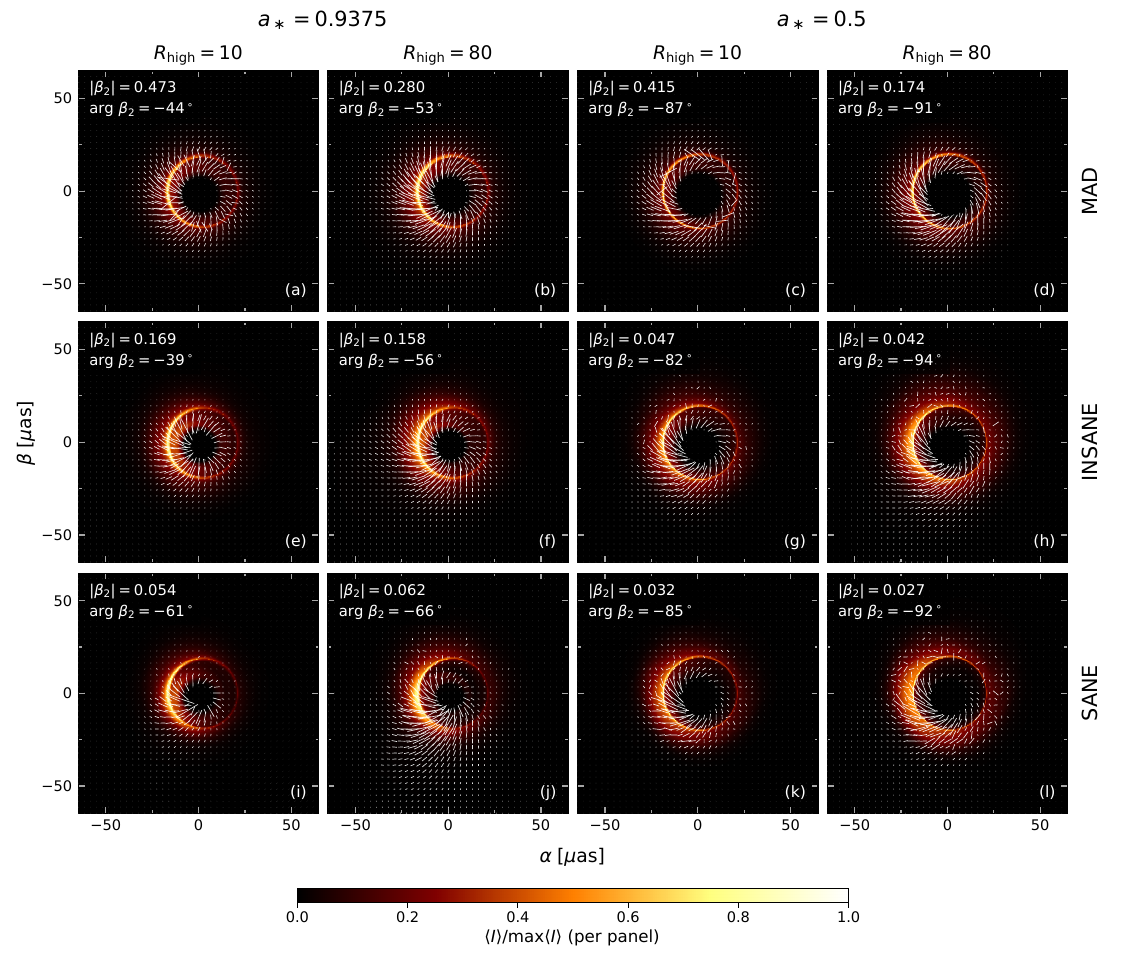}
\caption{Averaged intensity images with EVPA patterns between $t=15000\,M$ and $t=30000\,M$  at 230~GHz. In each panel, the GRRT images are produced by the post-processing of various GRMHD simulations at $a=0.94$ and $a=0.5$, and 
$R_{\rm high}=10$ and $R_{\rm high}=80$. Polarization ticks are displayed only in regions where $I>0.1\,I_\mathrm{max}$. The eDF used here are thermal distributions, and all images are produced from the total region of GRMHD simulations.}
\label{fig:I}
\end{figure*}

\cite{Palumbo2020} proposed a way to quantify the polarization morphology using the Fourier mode coefficients $\beta_m$:
\begin{align}
\beta_m & 
=\frac{1}{I_{\operatorname{ann}}}
\int_{\rho_{\min }}^{\rho_{\max }}
 \mathrm{d} \rho  \; \rho 
\int_0^{2 \pi} \mathrm{~d} \varphi \  P(\rho, \varphi) e^{-i m \varphi}, \\
I_{\operatorname{ann}} & 
=\int_{\rho_{\min }}^{\rho_{\max }} \mathrm{d} \rho\; \! \rho 
\int_0^{2 \pi} \mathrm{~d} \varphi \
  I(\rho, \varphi),
\end{align}
where $P(\rho, \varphi)$ and $I(\rho, \varphi)$ are the complex linear polarization and intensity located at the polar image-plane coordinates $(\rho, \varphi)$. The magnitude $|\beta_2|$ quantifies the strength of the rotationally symmetric polarization, while the phase $\angle \beta_2$ specifies the orientation of this symmetry relative to a purely radial EVPA pattern. For example, $\angle \beta_2=0$ corresponds to a purely radial polarization field, $\angle \beta_2 = \pi$ to a purely azimuthal field, and negative or positive $\angle \beta_2$ values indicate spirals with left- or right-handed helicity. Other modes $\beta_m,  m \neq 2$ represent the rotation asymmetry of the EVPA patterns.

Figure~\ref{fig:I} presents the time-averaged polarized images between 
$t=15000$~M and 
$30000$~M for the three accretion states, two black-hole spins, and two electron-heating prescriptions. We note that the torus sizes have comparable distributions of polarization properties, as shown in Fig.~\ref{fig:hist}, therefore, we focus on comparing the simulations using a large torus with $R_{\rm in}=20R_g$ for time-averaged properties. The EVPA morphology immediately reveals systematic differences between the magnetic configurations. As expected, the MAD models exhibit the most ordered polarization structure, with predominantly azimuthal EVPA patterns characteristic of highly organized magnetic fields. In contrast, the SANE models display noticeably more distorted EVPA structures, reflecting a greater degree of rotational asymmetry. The INSANE models occupy an intermediate regime, producing polarization patterns that are more ordered than those of SANE while remaining less symmetric than those of MAD. These qualitative differences are quantified by the quadrupolar polarization mode 
$\beta_2$, whose amplitude measures the strength of the rotationally symmetric polarization component. Across all combinations of black-hole spin and $R_{\rm high}$, the INSANE models yield intermediate 
$|\beta_2|$ values between the MAD and SANE limits, confirming that their average polarization morphology is likewise intermediate. The dependence on the electron-heating prescription is comparatively weak for the INSANE and SANE models, whereas the MAD models exhibit a stronger response to changes in 
$R_{\rm high}$. Increasing the black-hole spin generally produces more symmetric EVPA patterns, although the overall ordering of the three accretion states remains unchanged.

\begin{figure*}
\centering
\includegraphics[width=2.01\columnwidth]{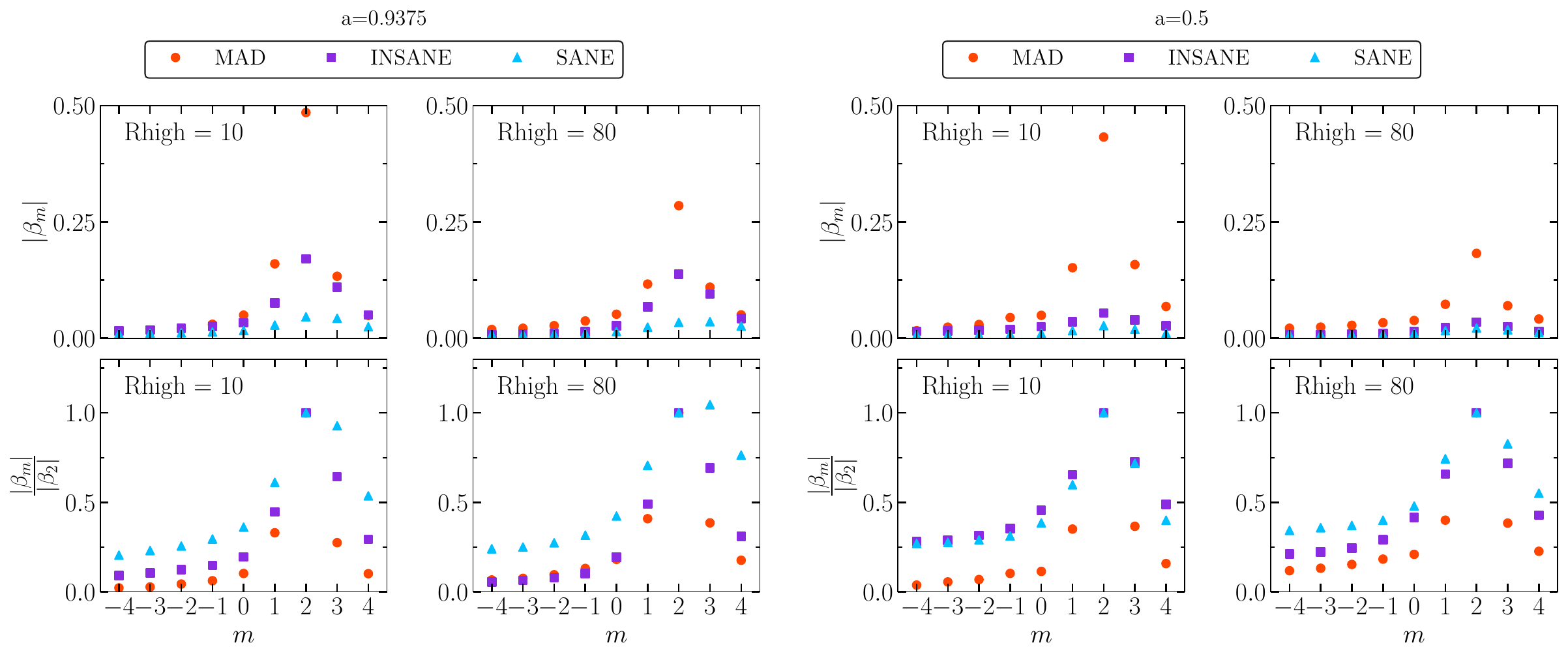}
\caption{Different beta modes of MAD, INSANE, and SANE models take the median value from $t=15000$~M to 30000~M. The top panel and bottom panel show $\beta_m$ and the ratio of $|\beta_m| / |\beta_2|$ at different modes for total emissions.}
\label{fig:beta_m}
\end{figure*}

The average polarization morphology can be quantified further by examining the complete spectrum of Fourier modes. Figure~\ref{fig:beta_m} shows the amplitudes of the polarization coefficients averaged over 
$t=15000$~M and $30000$~M. The upper panel presents the absolute amplitudes $|\beta_m|$. The overall polarization strength decreases systematically from MAD to INSANE and finally to SANE, consistent with the increasing level of magnetic disorder seen in the polarization images. To isolate the degree of rotational asymmetry from the overall polarization amplitude, the lower panel shows the normalized mode ratios 
$|\beta_m|/|\beta_2|$, following \citet{Long2026}. Here, $|\beta_2|$
 represents the dominant rotationally symmetric component expected from ring-like emission, while modes with $m \neq 2$ quantify deviations from this symmetry. The SANE models consistently exhibit the largest relative power in the asymmetric modes, demonstrating that their polarization morphology is intrinsically less axisymmetric than that of the other accretion states. For the 
high spin $a=0.9375$ case, the 
$m=3$ component is comparable to or even exceeds the quadrupolar mode ($|\beta_3|/|\beta_2|\simeq0.93$ and 1.05 for $R_{\rm high}=10$ and 80, respectively), indicating that asymmetric structures dominate the polarization morphology. At the opposite extreme, the MAD models retain only weak higher-order modes, consistent with their highly ordered magnetic fields. The INSANE models again occupy an intermediate regime, with asymmetric mode amplitudes lying between the MAD and SANE limits.

\begin{figure*}
\centering
\includegraphics[width=2\columnwidth]{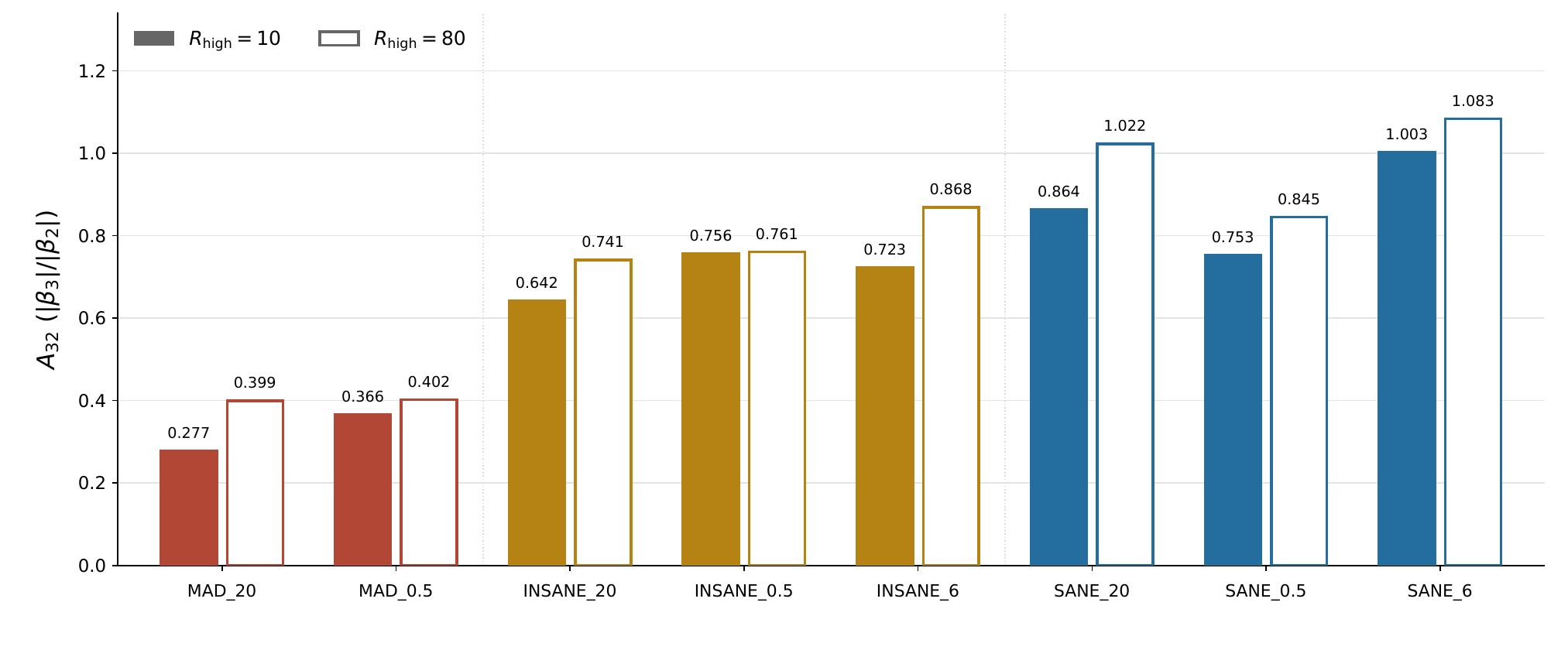}
\caption{Relative amplitudes of the third and second azimuthal polarization modes for the MAD (red), INSANE (gold), and SANE (blue) models. Bars show $A_{32}=\operatorname{median}_{t}\!\left(|\beta_3(t)|/|\beta_2(t)|\right)$, evaluated over $15000$--$30000\,M$. Filled and outlined bars represent $R_{\rm high}=10$ and $80$, respectively.  Larger $A_{32}$ indicates a stronger third-mode contribution relative to the second mode; MAD models exhibit systematically lower ratios across the models.}
\label{fig:a32}
\end{figure*}

We define $A_{32}=\operatorname{median}_{t}\!\left(|\beta_3(t)|/|\beta_2(t)|\right)$, as a characteristic of polarization asymmetry. Figure~\ref{fig:a32} shows that MAD models consistently yield the lowest values, $A_{32}\simeq0.28$--$0.40$, compared with $0.64$--$0.87$ for INSANE and $0.75$--$1.08$ for SANE. The systematically smaller MAD ratios indicate a stronger relative contribution from the rotationally symmetric component across the spins and electron-heating prescriptions explored here. At matched spin, initial torus size, and $R_{\rm high}$, INSANE generally lies between MAD and SANE, although the separation between INSANE and SANE depends on the model parameters. In particular, their ratios nearly coincide at $a=0.5$ and $R_{\rm high}=10$, with $A_{32}\simeq0.75$ for both states. By contrast, the high-spin SANE models approach or exceed unity, indicating that the third-mode amplitude is typically comparable to or larger than that of the second mode. Thus, $A_{32}$ captures a systematic change in the polarization morphology across the accretion states and is particularly useful for distinguishing MAD from INSANE and SANE within our model set.


\section{Multi-scale variability of intensity}
\label{sec:varibility}


In this section, we investigate the variability of the intensity time series.
Figure~\ref{fig:lc_I} shows the total intensity light curves, which exhibit variability on both short and long timescales. The variability amplitude and secular evolution depend on the accretion state and initial torus configuration: models initialized with a larger torus ($r_{\rm in}=20\,R_g$) display pronounced long-term trends, whereas the small-torus models show weaker secular evolution.

The fractional variability, $\sigma/\mu$ \citep{EHTC2022_SgrA_V,Wielgus2022}, summarizes the overall fluctuation amplitude but does not distinguish between timescales or reveal how variability evolves with time. Moreover, its measured value depends on the duration of the light curve \citep{Wielgus2022} and secular trends can dominate the variance and obscure the value. We therefore use time-frequency analysis to characterize the evolving variability of the intensity, together with the ratio of high- to low-frequency power as a diagnostic of changes in the relative contributions of rapid and slow fluctuations.

\subsection{Time–frequency analysis}

The small-torus INSANE\_6 and SANE\_6 simulations exhibit substantially weaker long-term evolution and therefore provide a more representative basis for time-series analysis. We therefore focus on the analysis of the small torus in this subsection, and similar analysis for other models can be found in the Appendix~\ref{app:time-analysis}.

To characterize the variability as a function of both time and frequency, we analyse the 230\,GHz total intensity $I$ with two complementary methods, the continuous wavelet transform (CWT) and the Hilbert--Huang transform (HHT). The CWT projects the light curve onto a fixed basis of scaled wavelets, whereas the HHT decomposes it adaptively into intrinsic mode functions (IMFs), each with a time-dependent instantaneous frequency and amplitude, and is well suited to non-linear and non-stationary signals. Features recovered by both methods are therefore unlikely to be artefacts of either decomposition. To limit edge effects, we mask the part of each scaleogram that lies below the cone of influence (COI) and discard the first and last 10 per cent of each Hilbert spectrum. The COI also sets the lowest accessible frequency, which is $\simeq 1.9\times10^{-4}\,M^{-1}$ for the full interval and $\simeq 5.7\times10^{-4}\,M^{-1}$ for the shorter interval considered below.

\begin{figure*}
\centering
\includegraphics[width=2.0\columnwidth]{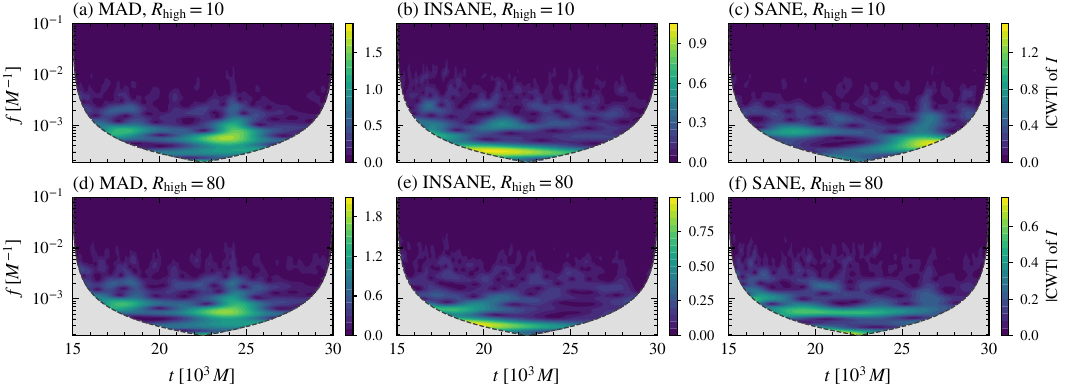}
\includegraphics[width=2.0\columnwidth]{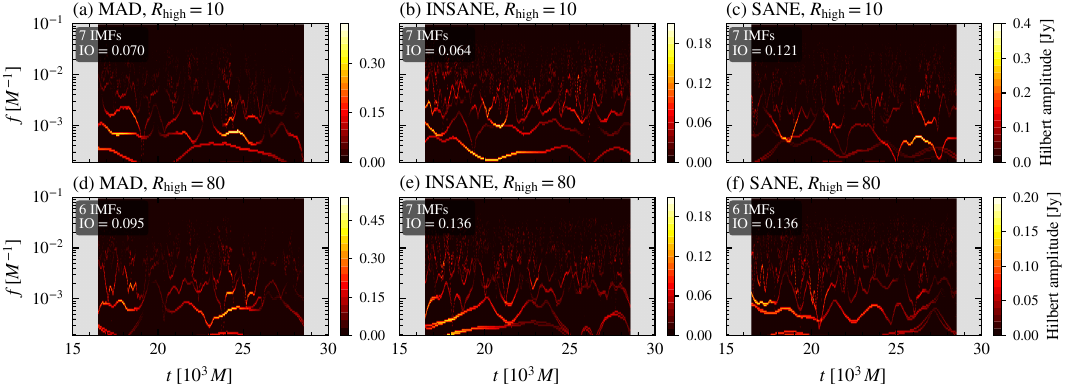}
\caption{Time--frequency analysis of the 
computed 230\,GHz intensity light
  curves, 
  over $15\,000$--$30\,000\,M$ for the MAD, INSANE and SANE
  models (columns) with $R_{\rm high}=10$ and $R_{\rm high}=80$ (rows).
  \emph{Top:} continuous wavelet transform scaleograms, computed with a
  complex Morlet wavelet. The colour scale shows the wavelet amplitude
  $|{\rm CWT}|$. The dashed curve marks the cone of influence; the grey region
  below it is affected by edge effects and is excluded.
  \emph{Bottom:} Hilbert--Huang spectra of the same light curves, obtained by
  empirical mode decomposition followed by the Hilbert transform of each
  intrinsic mode function (IMF). The colour scale shows the instantaneous
  Hilbert amplitude in Jy. The grey strips mask the first and last 10 per cent
  of the time range, where the decomposition is unreliable. The number of IMFs
  and the index of orthogonality (IO) are given in each panel.
  In both sets of panels, time is in units of $10^3\,M$ and frequency in
  $M^{-1}$, and each panel has its own colour scale. 
}
\label{fig:I_0.9375_co}
\end{figure*}

Fig.~\ref{fig:I_0.9375_co} shows the analysis over $t=15\,000$--$30\,000\,M$. In the MAD model, the wavelet amplitude is concentrated in two episodes, at $t\approx17\,000$--$18\,500\,M$ and $t\approx23\,000$--$25\,000\,M$, with characteristic frequencies $f\approx(5$--$8)\times10^{-4}\,M^{-1}$, i.e.\ time-scales of $\approx1300$--$1800\,M$. Each is associated with a large flare of the light curve that ends in a rapid decline. In the second episode $I$ rises from $\approx0.6$ to $\approx1.5$\,Jy over $\approx1000\,M$ and then falls by an order of magnitude within $\approx400\,M$, at the time of the deepest drop of the horizon magnetic flux in this simulation ($t\approx24\,400$--$24\,900\,M$ in Fig.~\ref{fig:phi}). 
This supports an association with a magnetic-flux eruption. The Hilbert spectra show their brightest ridges at the same times and at $f\approx(0.3$--$1)\times10^{-3}\,M^{-1}$, and the instantaneous frequency changes by a factor of a few across each event, as expected for a transient rather than for a coherent oscillation. The MAD panels for $R_{\rm high}=10$ and $R_{\rm high}=80$ are almost indistinguishable.

The INSANE model shows no comparably localized episode. Its largest wavelet amplitudes lie at lower frequency, $f\approx3\times10^{-4}\,M^{-1}$ (a time-scale of $\approx3000\,M$), which is also where the time-averaged wavelet spectrum peaks for both values of $R_{\rm high}$. This component is strongest at $t\approx19\,500$--$21\,000\,M$. It remains visible over most of the interval inside the COI for $R_{\rm high}=10$ and fades after $t\approx22\,000\,M$ for $R_{\rm high}=80$. The HHT assigns it to one or two low-frequency IMFs whose instantaneous frequencies drift by a factor of 2--3 around this value. It reflects the broad maximum of the light curve at $t\approx17\,000$--$19\,000\,M$ and the subsequent slow decline. Only about five cycles of this period fit within the analysed interval, and the temporal resolution of the wavelet at this frequency is several thousand $M$, so we interpret it as a slow modulation and not as a quasi-periodic oscillation. At $f\gtrsim10^{-3}\,M^{-1}$ the INSANE scaleograms contain many weak, short-lived features that are distributed evenly in time.

The SANE model is the most sensitive to the electron-heating prescription. For $R_{\rm high}=10$ the scaleogram is dominated by one feature at $t\approx26\,000$--$27\,500\,M$ and $f\approx5\times10^{-4}\,M^{-1}$ (a time-scale of $\approx2000\,M$), produced by a dip of $I$ to $\approx0.2$\,Jy followed by a flare to $\approx1$\,Jy at $t\approx26\,400\,M$. The Hilbert amplitudes of four of the seven IMFs peak within this interval. For $R_{\rm high}=80$ the same flare reaches only $\approx0.6$\,Jy, and the scaleogram instead shows a faint band at $f\approx(5$--$6)\times10^{-4}\,M^{-1}$ of nearly constant amplitude between $t\approx17\,000$ and $27\,000\,M$. The HHT recovers this band as an IMF with a median frequency of $6\times10^{-4}\,M^{-1}$ and an amplitude of $\approx0.08$\,Jy, superposed on a secular decline of the flux that accounts for the power at the lowest resolved frequencies. On time-scales longer than $\sim1000\,M$, the intensity variability of the MAD model and of the SANE model with $R_{\rm high}=10$ is therefore dominated by a few discrete, large-amplitude events, whereas that of the INSANE model is a slow modulation without such events.

\begin{figure*}
\centering
\includegraphics[width=2.0\columnwidth]{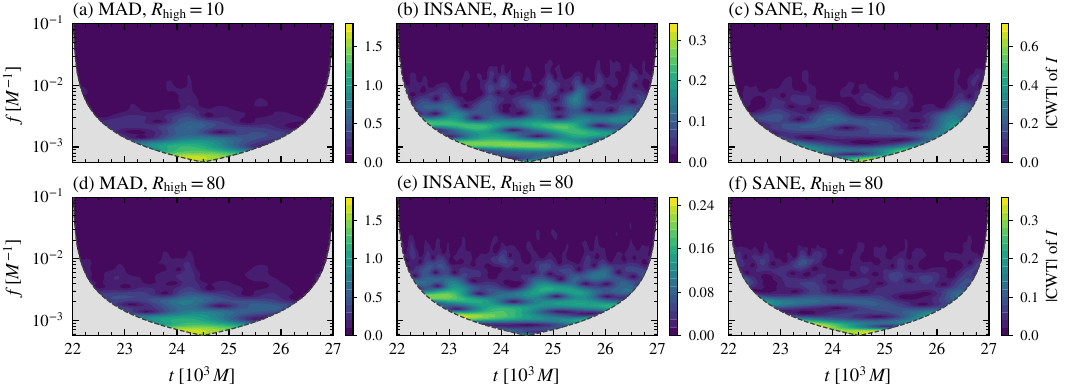}
\includegraphics[width=2.0\columnwidth]{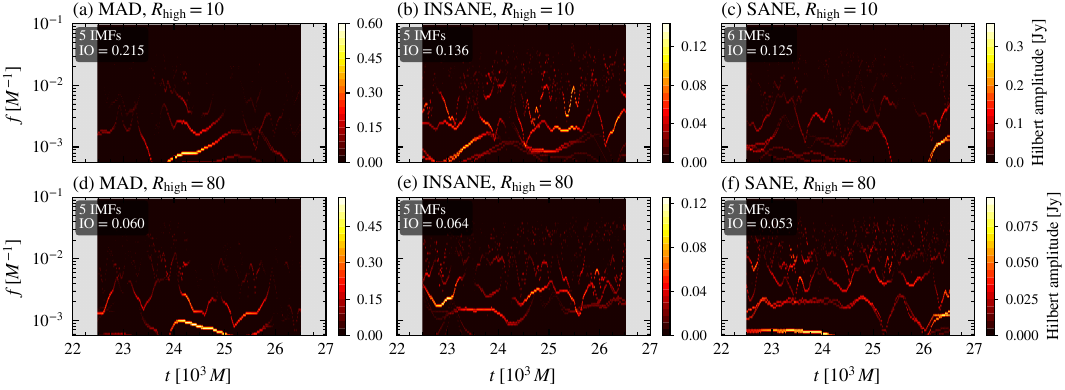}
\caption{ As Fig.~\ref{fig:I_0.9375_co}, but for the interval
  $22\,000$--$27\,000\,M$ of the same light curves. 
}
\label{fig:I_0.9375_short}
\end{figure*}

To examine shorter time-scales, Fig.~\ref{fig:I_0.9375_short} repeats the analysis of the same light curves over $t=22\,000$--$27\,000\,M$. Restricting the interval removes the slow components described above, and the three accretion states then differ more clearly. The MAD panels contain a single feature, centred on the flux collapse at $t\approx24\,100$--$24\,400\,M$, which appears in the scaleogram as a narrow streak extending from the lowest accessible frequency to $f\sim10^{-2}\,M^{-1}$. The Hilbert amplitudes of nearly all IMFs peak at this time, and one IMF with a median frequency of $\approx10^{-3}\,M^{-1}$ carries 50--63 per cent of the total IMF energy in the unmasked part of the interval.

The INSANE light curve contains no dominant event in this interval. Its fractional rms variability is 14--21 per cent, compared with $\approx58$ per cent for the MAD model and 21--33 per cent for the SANE model. Its wavelet amplitude is instead spread over most of a decade in frequency. It exceeds half of its maximum up to $f\approx(3$--$4)\times10^{-3}\,M^{-1}$, compared with $\approx10^{-3}\,M^{-1}$ for the MAD model and $(2$--$3)\times10^{-3}\,M^{-1}$ for the SANE model, and it consists of several local maxima of comparable strength, with periods between $\approx300$ and $\approx1800\,M$, that are spread over the whole interval. The HHT gives the same picture. The energy is shared among several IMFs, and the most energetic one in the unmasked part has a median frequency of $(2.4$--$3.1)\times10^{-3}\,M^{-1}$ (a time-scale of $\approx320$--$410\,M$), a factor of $\approx3$ higher than the $(6$--$13)\times10^{-4}\,M^{-1}$ found for the MAD and SANE models. This behaviour is consistent with stochastic, broad-band variability without a preferred time-scale. In absolute terms, however, the INSANE wavelet amplitude at $f\gtrsim10^{-3}\,M^{-1}$ is comparable to that of the SANE model and lower than that of the MAD model. The INSANE model is distinguished by the absence of a dominant low-frequency event, and not by an excess of short-time-scale power.

The SANE model is intermediate. Its amplitude is concentrated at the lowest accessible frequencies, with weak and fragmented power above $10^{-3}\,M^{-1}$. For $R_{\rm high}=10$ the amplitude increases towards the end of the interval, at the onset of the flare at $t\approx26\,400\,M$ identified above, which lies close to the region affected by edge effects. The CWT and the HHT recover the same ordering of the three accretion states for both values of $R_{\rm high}$.

\subsection{Fast power fraction}

The time--frequency maps of the previous subsection suggest that the accretion states differ in how their variability is distributed over time-scales, and not only in its amplitude. To quantify this for all models, we define the fast power fraction,
\begin{equation}
    F_{\mathrm{fast},I} =
    \frac{\overline{E}_{I,[P_0,P_{\rm f}]}}{\overline{E}_{I,[P_0,P_{\rm b}]}} \, ,
    \label{eq:Ffast}
\end{equation}
where $E_{I,[P_1,P_2]}(t)$ is the instantaneous variability energy of the total-intensity light curve in the period band $P_1$--$P_2$, the overline denotes a time average, and $[P_0,P_{\rm f}]$ and $[P_0,P_{\rm b}]$ are the fast and the broad band, respectively, with $P_{\rm f}<P_{\rm b}$. The numerator and the denominator are averaged over the same time samples. $F_{\mathrm{fast},I}$ is therefore the fraction of the variability energy at periods $P_0$--$P_{\rm b}$ that is carried by periods shorter than $P_{\rm f}$. It describes the shape of the variability spectrum, is independent of the flux normalization of the light curve, and is not a measure of the variability amplitude.

We evaluate the instantaneous variability energy with both methods. For the CWT, 
\begin{equation}
    E^{\rm CWT}_{I,[P_1,P_2]}(t) =
    \int_{\ln P_1}^{\ln P_2} \frac{|W(s,t)|^2}{s}\,{\rm d}\ln P \, ,
    \label{eq:Ecwt}
\end{equation}
where $W(s,t)$ is the wavelet coefficient of the mean-subtracted light curve at scale $s$, and $P\simeq1.03\,s$ is the equivalent Fourier period of the Morlet wavelet with $\omega_0=6$. This is the usual scale-averaged wavelet power, up to a constant that cancels in equation~(\ref{eq:Ffast}). For the HHT,
\begin{equation}
    E^{\rm HHT}_{I,[P_1,P_2]}(t) =
    \frac{1}{2}\sum_{k} a_k^2(t)\,\Theta_k(t) \, ,
    \label{eq:Ehht}
\end{equation}
where $a_k(t)$ and $f_k(t)$ are the instantaneous amplitude and frequency of the $k$th IMF, and $\Theta_k(t)=1$ if $1/P_2\le f_k(t)<1/P_1$ and $\Theta_k(t)=0$ otherwise. The factor $1/2$ makes the contribution of a sinusoidal mode equal to its variance. Times at which no IMF lies in the band contribute zero energy to the average. The two estimators weight the signal differently, so their values of $F_{\mathrm{fast},I}$ need not coincide.

We adopt $P_0=80\,M$ and $P_{\rm b}=1500\,M$. The lower limit corresponds to eight samples per period at the cadence of $10\,M$. The upper limit is set by edge effects: after removing $2730\,M$ at each end of the light curves, a conservative allowance for the cone of influence at $P=1500\,M$, the retained interval $17\,730$--$27\,270\,M$ still covers 6.4 periods of the slowest component. The boundary $P_{\rm f}$ is the only free choice. For a circular Keplerian orbit, $P=2\pi(r^{3/2}+a)\,M$, periods of 80, 200 and $1500\,M$ correspond to radii of $r\approx5$, 10 and $38\,r_{\rm g}$ for both spins considered here. A boundary of a few hundred $M$ therefore separates the orbital time-scales of the inner $\approx10\,r_{\rm g}$ from those of the flow at larger radii, and it is also of the order of the time-scales expected for turbulent fluctuations in that region. We do not attempt to identify $P_{\rm f}$ with a specific physical time-scale in this work. We present results for $P_{\rm f}=200\,M$ as an illustrative choice, and we have repeated the measurement for every boundary between 200 and $500\,M$ in steps of $1\,M$.

\begin{figure*}
\centering
\includegraphics[width=2\columnwidth]{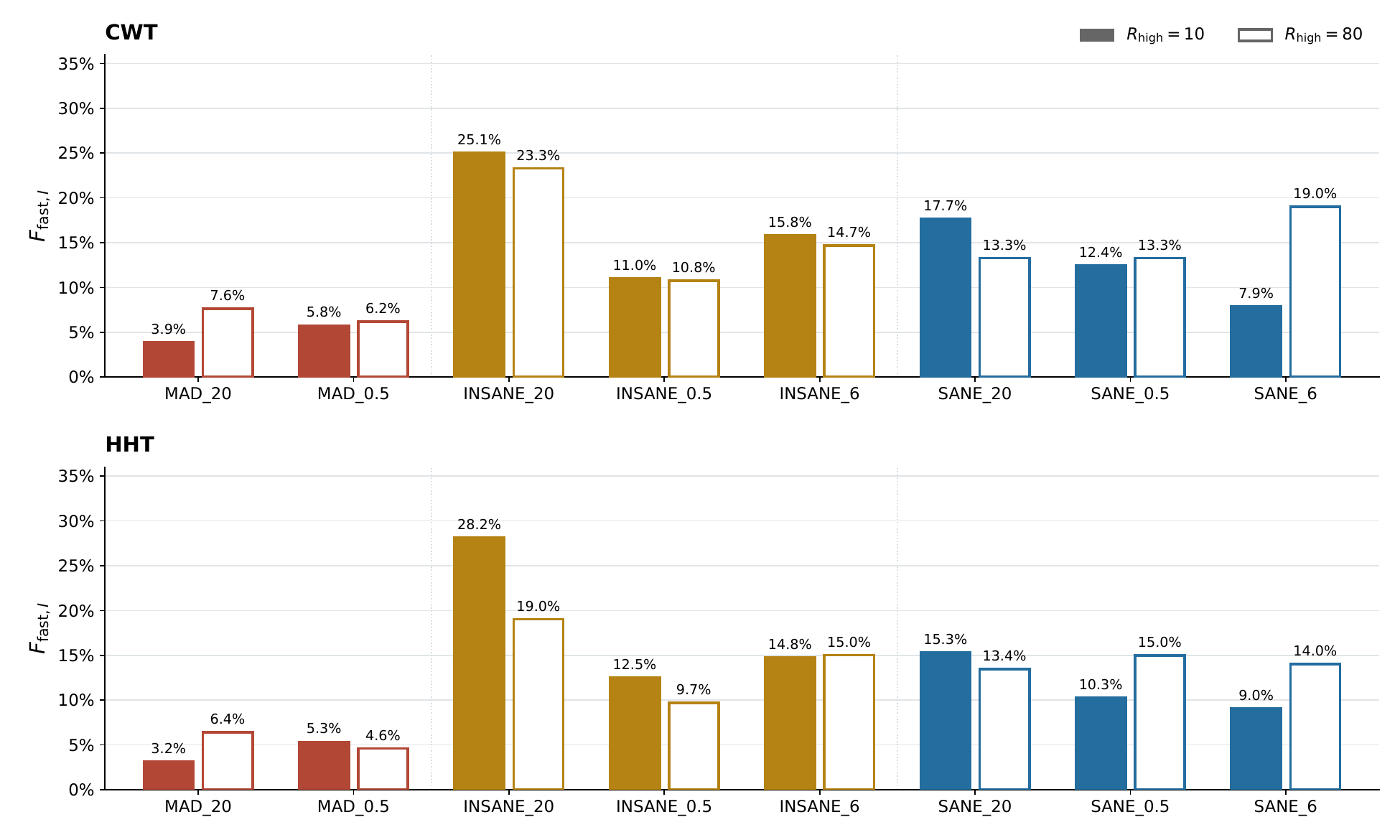}
\caption{Fraction of 
the 230-GHz intensity variability,  
in terms of fast  power fraction $F_{\mathrm{fast},I}$ (equation~\ref{eq:Ffast}), for all 16
  models, 
  computed from continuous wavelet transform (CWT; top) and 
  Hilbert--Huang transform (HHT; bottom). The fast band is $80$--$200\,M$ and
  the broad band is $80$--$1500\,M$. The variability energies are averaged over the time interval 
  $17\,730$--$27\,270\,M$, which excludes the parts of the
  $15\,000$--$30\,000\,M$ light curves affected by edge effects at a period of
  $1500\,M$. Colours denote the accretion state (red: MAD; yellow: INSANE;
  blue: SANE), and filled and open bars show $R_{\rm high}=10$ and
  $R_{\rm high}=80$, respectively. 
  The
  R20 and Rin6 models have black hole spin $a=0.9375$ and the $0.5$ models have
  $a=0.5$; no MAD\_6 run is available. 
Note that the values are fractions of
  variability energy, not fractional rms amplitudes. Both panels share the
  same vertical scale.}
\label{fig:Ffast}
\end{figure*}

Fig.~\ref{fig:Ffast} shows $F_{\mathrm{fast},I}$ for all 16 models. The MAD models have the lowest values, 3.9--7.6 per cent with the CWT and 3.2--6.4 per cent with the HHT, compared with 10.8--25.1 and 9.7--28.2 per cent for the INSANE models and 7.9--19.0 and 9.0--15.3 per cent for the SANE models. The MAD range does not overlap that of either of the other states. At fixed spin, torus size and $R_{\rm high}$, the MAD value is lower than the INSANE value by a factor of 1.7--8.9 and lower than the SANE value by a factor of 1.7--4.9. The two methods agree to within 5 percentage points for every model, with a median difference of 1.3 percentage points. A low $F_{\mathrm{fast},I}$ does not imply weak variability. In the previous subsection, the MAD light curve was found to be dominated by a few large-amplitude events with time-scales $\gtrsim1000\,M$, which we associated with magnetic-flux eruptions. Such events place most of the energy at the long-period end of the broad band and thereby lower the fast fraction. The MAD values also depend little on the electron-heating prescription, changing by at most 4 percentage points between $R_{\rm high}=10$ and 80.

The INSANE and SANE models are less clearly separated. The medians over the models of each state follow the hierarchy INSANE $>$ SANE $>$ MAD, with 15, 13 and 6 per cent for the CWT and 15, 14 and 5 per cent for the HHT, and the highest fractions of all, 19--28 per cent, are reached by INSANE\_20. For the R20 models the INSANE value exceeds the SANE value by 6--13 percentage points. The ranges of the two states nevertheless overlap, and the hierarchy is not satisfied by every matched pair. For the $a=0.5$ models the two states are reversed with the CWT for both values of $R_{\rm high}$ (by 1.4 and 2.5 percentage points) and with the HHT for $R_{\rm high}=80$ (10 per cent for INSANE against 15 per cent for SANE), and for the Rin6 models with $R_{\rm high}=80$ they are reversed with the CWT (15 per cent for INSANE against 19 per cent for SANE) but not with the HHT (15 against 14 per cent). The dependence on $R_{\rm high}$ is also larger than in the MAD case for some of these models. The strongest example is SANE\_6, for which $F_{\mathrm{fast},I}$ increases from 8--9 per cent for $R_{\rm high}=10$ to 14--19 per cent for $R_{\rm high}=80$. This is consistent with the previous subsection, where the $R_{\rm high}=10$ light curve of this model was found to be dominated by a single slow dip and flare that is much weaker for $R_{\rm high}=80$.

These results are not specific to $P_{\rm f}=200\,M$. All fractions increase with $P_{\rm f}$, the state medians rising from 5--6 (MAD), 15 (INSANE) and 13--14 per cent (SANE) at $P_{\rm f}=200\,M$ to 24--28, 44--48 and 40--43 per cent at $500\,M$. The MAD value is the lowest in every matched comparison, with both methods, over the whole range $200$--$500\,M$, and the hierarchy of the state medians is preserved throughout, although the INSANE and SANE medians come within 1 percentage point of each other at $P_{\rm f}\approx290$--$340\,M$ with the CWT and at $\approx360$--$410\,M$ with the HHT. The exceptions concern only the ordering of the INSANE and SANE models. INSANE\_0.5 falls below SANE\_0.5 at all boundaries with the CWT for both values of $R_{\rm high}$ and with the HHT for $R_{\rm high}=80$, and at $P_{\rm f}\ge307\,M$ with the HHT for $R_{\rm high}=10$. INSANE\_6 falls below SANE\_6 for $R_{\rm high}=80$ at $P_{\rm f}\le319\,M$ with the CWT, and INSANE\_20 falls below SANE\_20 for $R_{\rm high}=80$ at $P_{\rm f}\ge361\,M$ with the HHT. The reversals reach at most 22 percentage points.

\section{Summary and Discussion}

By systematically comparing MAD, SANE, and INSANE GRMHD simulations coupled with polarized radiative transfer calculations, we identify two diagnostics that may help distinguish between different magnetization states and may provide useful probes of horizon-scale accretion flows such as M87$^\ast$.

Our analysis shows that 
\begin{enumerate}
    \item MAD model produces the most symmetric EVPA patterns, and SANE produces the least symmetric pattern, whereas INSANE lies between them.
    \item The multi-scale temporal behaviour and its variability distribution of the intensity differs qualitatively between the three states. 
While MAD is characterized by time-localized oscillatory variability associated with magnetic-flux eruptions and SANE shows weaker oscillatory behaviour, INSANE exhibits persistent broadband multi-frequency variability.
\end{enumerate}
Together, these results demonstrate that combining polarization morphology with temporal analysis provides more information than static imaging alone.

Recent EHT observations of M87$^\ast$ and Sgr~A$^\ast$ have demonstrated the importance of polarimetric measurements for distinguishing accretion-flow models \citep{EHTC2022_PolarizationV,EventHorizonTelescope:2025vum}.
The Fourier decomposition provides a framework for quantifying the polarization morphology \citep{Palumbo2020, Palumbo2022ApJ,Palumbo2023ApJ}, typically using the quadrupolar mode $\beta_2$, which characterizes the dominant rotationally symmetric component of the EVPA field. 
Our analysis shows the importance of considering modes  $\beta_m$  ($m \neq 2$) and the ratio of $|\beta_{m}| / |\beta_2|$ \citep{Long2026} as measures of deviations from the rotational symmetry. 
The retention of the $\beta_m$ in the Fourier decomposition for quantifying polarization morphology is essential for probing the magnetic field structure in the accretion flow. 
More specifically, we have identified the ratio $|\beta_3|/|\beta_2|$ showing the clearest separation between the three accretion states, whereas solely considering $|\beta_1|$, $|\beta_2|$ \citep[see e.g., in][]{EventHorizonTelescope:2025vum} and their ratio is inadequate to fully capture the magnetic structure information. 
Our results, therefore, suggest that future multi-epoch EHT observations could use the temporal $|\beta_3|/|\beta_2|$ as one of the sensitive observational probes of the magnetic structure of horizon-scale accretion flows.

The temporal and polarization properties of horizon-scale emission carry important information about the magnetic flux properties of the accretion flow. 
When future EHT-like observations improve in cadence and sensitivity, temporal measurements will become an increasingly important observational diagnostic.  
The different temporal signatures identified in this work likely reflect the distinct dynamical processes governing the accretion states in the three models. 
In the timing analysis of the MAD simulations, both the Continuous Wavelet Transform and Hilbert--Huang Transform identify localized oscillatory features associated with magnetic-flux eruption events consistently. 
The eruptions are a characteristic of strongly magnetized accretion flows, in which the accumulation of magnetic flux periodically disrupts the inflow before the excess magnetic flux is expelled. 
The localized oscillatory signatures, therefore, likely trace the dynamical reconfiguration of the inner accretion flow during these eruptive episodes. 
The SANE simulations exhibit weaker localized oscillatory behaviour, and the presence of identifiable localized oscillatory features depends on the adopted electron-heating prescription. This sensitivity suggests that, in weakly magnetized disks, the observed temporal variability depends not only on the underlying MHD turbulence but also on the emission region of synchrotron radiation. 
The INSANE simulations display a qualitatively different form of variability. Instead of isolated oscillatory events, they exhibit persistent broadband multi-frequency behaviour spanning a wide range of timescales. This behaviour is consistent with variability arising from fluctuations over multiple spatial and temporal scales, without a single dominant oscillatory component. A quantifiable measurement of the variability distribution across temporal scales, $F_{\mathrm{fast},I}$, identifies a hierarchy of INSANE $>$ SANE $>$ MAD for each radiative model.  The agreement between the Continuous Wavelet Transform and Hilbert--Huang Transform suggests that the broadband variability is unlikely to be an artifact of a particular time--frequency decomposition. Although both methods remain susceptible to finite-duration effects and non-stationarity, they rely on fundamentally different approaches to signal decomposition, making it less likely that the observed broadband structure is produced by the analysis itself.

The analysis above for diagnosing the observables is particularly useful for the ongoing investigation of alternative magnetisation states of accretion flows in GRMHD simulations. 
Traditionally, black-hole accretion has largely been framed in terms of two limiting magnetisation states in terms of the SANE and MAD models \citep{Narayan2012}. These two states are distinguished in GRMHD simulations, typically using physical quantities 
such as the magnetic flux, the mass accretion rate, the jet efficiency, and the dynamical influence of the magnetic field. 
However, recent numerical studies \citep[e.g.][]{Raha2026MNRAS, wong2026sanemadchimera} suggested that magnetized accretion flows occupy a broader range of physical states than this simple dichotomy. 
Intermediate magnetic-flux configurations, alternative magnetic-flux supply histories, and partially magnetically arrested accretion flows have all been shown to exhibit properties that cannot be fully characterized within the conventional framework in terms of the MAD/SANE model \citep{Ressler2020ApJ, Manikantan2024ApJ, Raha2026MNRAS}. 
All these developments motivate the search for observable diagnostics capable of distinguishing such accretion states. 

Our analysis addresses this need by identifying observable time-resolved polarization quantities that distinguish different magnetised accretion disk states. While the time-averaged polarization morphology places INSANE naturally between the MAD and SANE, its temporal behaviour of intensity and polarization is qualitatively distinct. 
In particular, the persistent broadband multi-frequency variability of intensity shown in both the Continuous Wavelet Transform and Hilbert--Huang Transform is absent from either MAD or SANE. 
These results demonstrate that temporal behaviour carries information beyond that contained in static images alone, providing an additional observational dimension for identifying partially arrested accretion flows.
Such diagnostics are well-suited to future time-domain EHT-like observations of M87$^\ast$, where long-term monitoring will allow the temporal evolution of polarization morphology and variability to be measured directly. 
They may also prove valuable for Sgr~A$^\ast$, whose observed variability has proven difficult to reconcile with either canonical MAD or SANE models alone \citep{EventHorizonTelescope:2021bee}. 
Although comparisons with observations should ultimately include propagation effects such as external Faraday rotation, their effect on the temporal polarization diagnostics explored here is unknown.

Note that in this study, we consider only thermal electron distributions, a limited set of black-hole spins and electron-heating prescriptions, and a viewing geometry appropriate for M87$^\ast$. 
A more comprehensive work should consider a broader range of plasma models and magnetic-field configurations, and include non-thermal electron populations and various black-hole spins. 
Applying the same methodology to simulations for parameters appropriate for Sgr~A$^\ast$ will help verify whether partially arrested accretion flows can simultaneously reproduce its observed variability and polarization properties. 
More generally, as an increasingly diverse range of magnetized accretion states emerges, continued exploration via GRMHD simulations, combined with polarisation morphology analysis and multi-scale temporal analysis, offers a promising route for connecting horizon-scale observations and black-hole accretion physics.

\section*{Acknowledgements}  
We thank Dr Alvina On (NCTS) for valuable discussions on time series analysis and Dr Kaye Jiale Li (UCL) 
 on general aspects of the science.  
JSL is supported by UCL 
  through a UCL-RES scholarship 
  and by the TDLI visiting student program.  
AU acknowledges the support from the Research Fund for Excellent International PhD Students (grant No. W2442004).  
YM is supported by the National Key R\&D Program of China (grant No. 2023YFE0101200), the National Natural Science Foundation of China (grant Nos. 12273022, 12511540053, and W2641030), and the Shanghai municipality orientation program of basic research for international scientists (grant No. 22JC1410600).
KW is supported by a UKRI-STFC SA grant awarded 
  to UCL-MSSL. 
JSL and KW acknowledge the support from the UCL Cosmoparticle Initiative. 
The simulations were conducted 
  on the UCL-MSSL cluster 
  and the Siyuan Mark-I HPC 
  at Shanghai Jiao Tong University. 
This work has made use 
  of the NASA Astrophysics Data System. 
We acknowledge using ChatGPT 
  in polishing some of the sentences in the text.

\section*{Data Availability}

The data underlying this article will be shared on reasonable request to the corresponding author.



\bibliographystyle{mnras}
\bibliography{example} 




\appendix

\section{Details on the GRMHD simulation setups}
\label{app: GRMHD}

The initial magnetic configuration is prescribed through a single dipolar loop with vector potential 
\begin{align}
A_\phi^{\rm SANE} &=\max(q_{\rm S},0),\quad
q_{\rm S}=\rho/\rho_{\max}-0.2,\nonumber\\
A_\phi^{\rm INSANE} &=\max(q_{\rm S}^2r^3,0),\nonumber\\
A_\phi^{\rm MAD} &=\max(q_{\rm M},0),\nonumber\\
q_{\rm M}&=\frac{\rho}{\rho_{\max}}
\left(\frac{r}{r_{\rm in}}\right)^3
\sin^3\theta\,e^{-r/400}-0.2,
\end{align}

and the four horizon quantities:
\begin{align}
\dot M &=-\int\rho u^r\sqrt{-g}\,\mathrm{d}\theta\,\mathrm{d}\phi,\nonumber\\
\Phi_{\rm BH}&=\frac12\int|B^r|\sqrt{-g}\,\mathrm{d}\theta\,\mathrm{d}\phi,
\quad\phi_{\rm BH}=\frac{\Phi_{\rm BH}}{\sqrt{\dot M}},\nonumber\\
\dot E&=\int T^r{}_t\sqrt{-g}\,\mathrm{d}\theta\,\mathrm{d}\phi,
\quad\eta_{\rm out}=\frac{\dot M-\dot E}{\dot M}.
\end{align}
We use $c=1$ and positive $\dot M$ and $\dot E$ for inward fluxes. The ideal-GRMHD stress--energy tensor is
\begin{align}
T^\mu{}_\nu={}&(\rho+u_{\rm g}+p+b^2)u^\mu u_\nu\nonumber\\
&+(p+b^2/2)\delta^\mu{}_\nu-b^\mu b_\nu,
\end{align}
where $u_{\rm g}$ is the internal-energy density, $p$ the gas pressure and $b^\mu$ the comoving magnetic-field four-vector. Thus, $\dot E$ contains matter and electromagnetic energy fluxes, but no radiative flux because the simulations are non-radiative.

\begin{figure*}
\centering
\includegraphics[width=2.0\columnwidth]{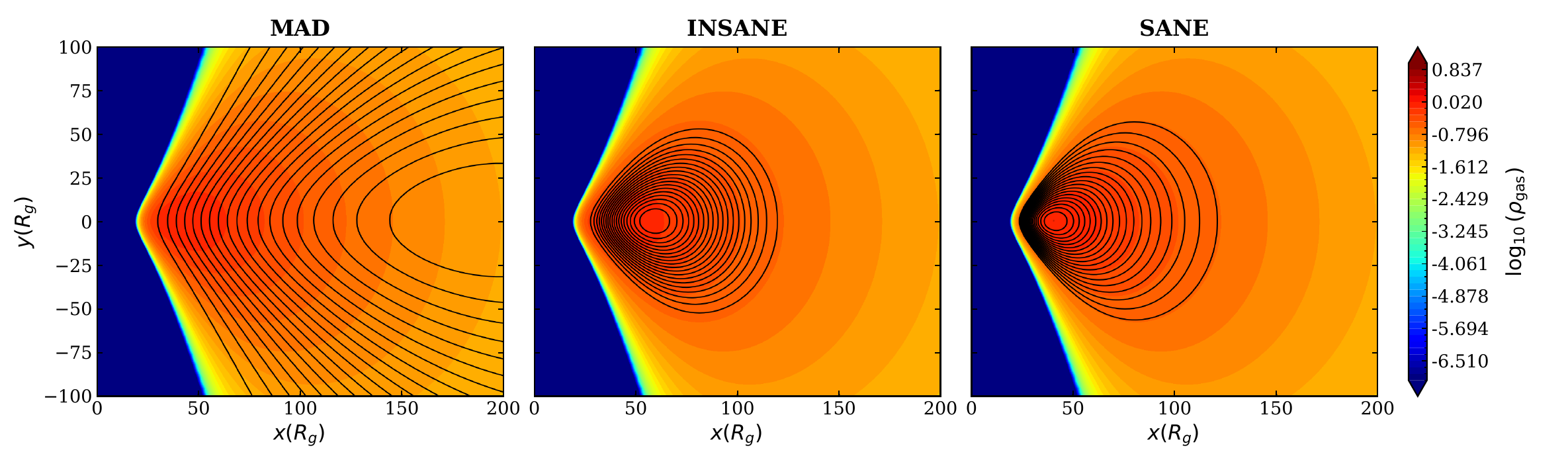}
\caption{(Updated figure with clarified figure captions) Initial configurations of MAD, INSANE, and SANE models (from left to right). For illustration purposes, we show the configurations for a torus with $r_{\rm in}=20R_g$.}
\label{fig:model}
\end{figure*}

In Fig.~\ref{fig:model}, we show the density profile of the initial torus configuration along with the corresponding magnetic field topologies for the MAD, INSANE, and SANE models. As mentioned previously, to ensure a consistent comparison among the different setups, we keep the size and structure of the initial torus identical for all three models with $r_{in}=20\,r_g$ and $r_{max}=41\,r_g$. For comparison with the commonly used configurations in the literature for the SANE model, we also consider an additional setup shown in the second row, where the torus parameters are $r_{in}=6\,r_g$ and $r_{max}=12\,r_g$ for the SANE and INSANE cases.

\begin{figure*}
\centering
\includegraphics[width=1.\columnwidth]{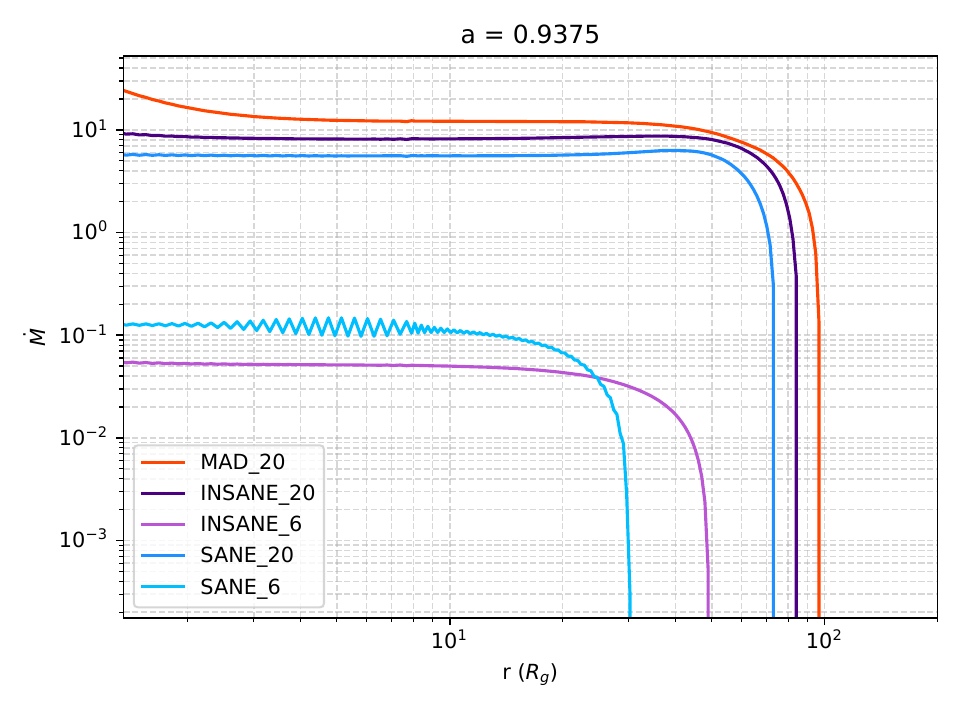}
\includegraphics[width=1.\columnwidth]{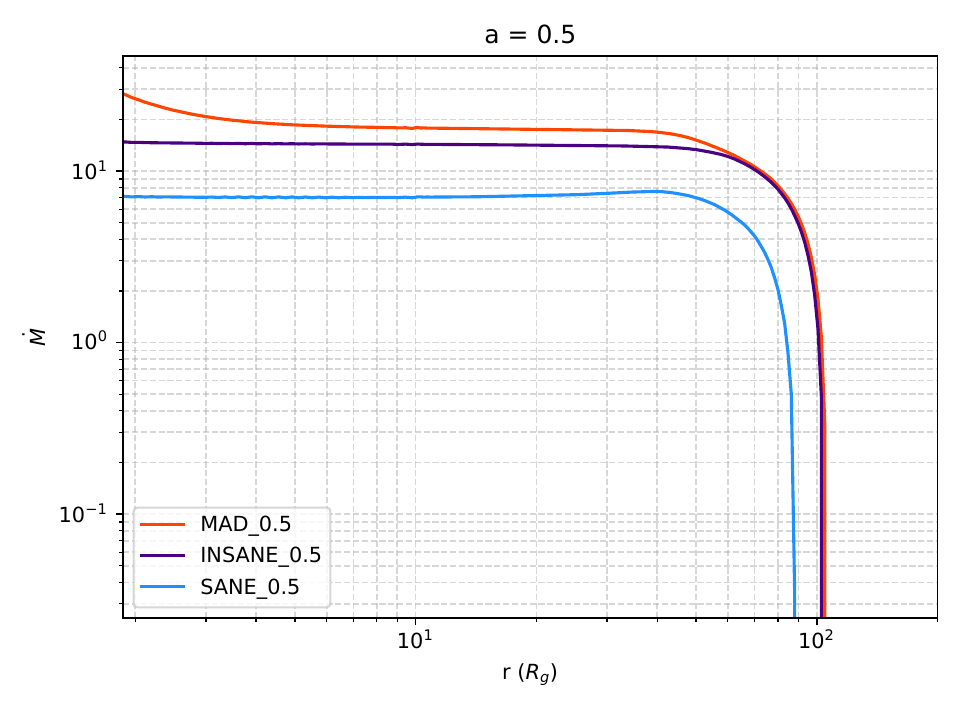}
\caption{Time-averaged radial profiles of the accretion rate $\dot{M}$ of the MAD, INSANE, and SANE models between 15000 M and 30000 M. The left panel has a spin of 0.9375, whereas the right has a spin of 0.5.}
\label{fig:validation}
\end{figure*}

To assess whether the radiative transfer calculations are performed on statistically steady accretion flows, Figure~\ref{fig:validation} shows the time-averaged radial profiles of the mass accretion rate $\dot{M}$, between 15000\,M and 30000\,M for all simulations. For the large-torus models, the accretion rate becomes approximately radius-independent within $r \lesssim 70 \ R_{\rm g}$, indicating that the inner accretion flow has reached a quasi-steady state despite the continuing evolution of the outer disc. For the small-torus models, $\dot{M}$ remains nearly constant within $r \lesssim 30\ R_{\rm g}$, demonstrating that the inflow equilibrium extends throughout the entire simulated domain.

The polarized radiative transfer calculations presented in this work are performed using emission within a radius of 17\,$R_{\rm g}$. This emitting region, therefore, lies well inside the quasi-steady inflow region for both the large- and small-torus simulations, ensuring that the synthetic images and light curves are not significantly affected by the non-equilibrium evolution of the outer accretion flow.

\section{Polarimetric properties}
\label{app:pol}

\begin{figure*}
\centering
\includegraphics[width=2.1\columnwidth]{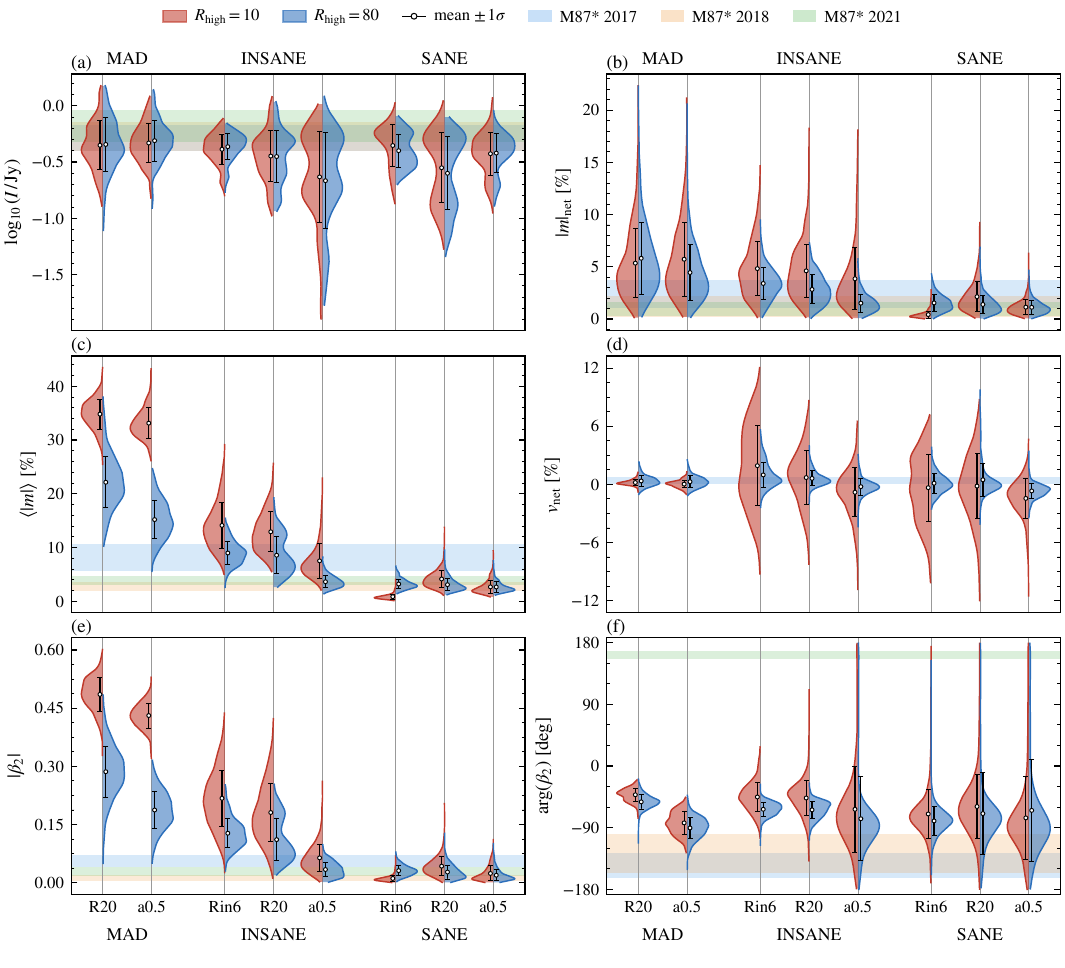}
\caption{Histograms of polarization and intensity quantities for the MAD, INSANE, and SANE models at 230~GHz. Results for $R_{\rm high}=10$ and $R_{\rm high}=80$ are shown in red and blue, respectively. From left to right, the columns correspond to MAD, INSANE, and SANE configurations. The shaded bands (light blue, light orange, and light green) indicate the ranges inferred from different years of EHT observations of M87 (2017, 2018, and 2021). The histograms between simulations using a large torus and a small torus at spin 0.9375 are similar.}
\label{fig:hist}
\end{figure*}

To further characterise the statistical properties of the simulated images, we analyse the polarization observables across the full ensemble of snapshots and present their distributions as histograms in Fig.~\ref{fig:hist}. The quantities considered include the amplitude and phase of the quadrupolar polarization mode, $|\beta_2|$ and $\arg(\beta_2)$; circular and linear polarization measures, $|v_{\rm net}|$, $\langle |m| \rangle$, and $|m_{\rm net}|$; as well as the total intensity $I$.

The MAD models generally produce higher linear polarization fractions compared with the INSANE and SANE models. In addition, the MAD configuration exhibits a stronger sensitivity to the electron heating parameter $R_{\rm high}$. In contrast, the INSANE and SANE models show comparatively weaker dependence of the linear polarization properties on the heating prescription.

For circular polarization, however, the behaviour is reversed. The INSANE and SANE models show noticeable differences in $|v_{\rm net}|$ between the two heating prescriptions ($R_{\rm high}=10$ and $R_{\rm high}=80$), whereas the MAD models produce similar levels of circular polarization for both heating choices. This suggests that circular polarization is more strongly influenced by plasma thermodynamics in weakly magnetized flows, while in strongly magnetized MAD configurations, the magnetic field geometry dominates the polarization structure.

\section{Complete Time-series analysis}
\label{app:time-analysis}

The main text focuses on the INSANE\_6 and SANE\_6 models because their weaker secular evolution makes them better suited for quantitative time-series analysis. For completeness, we present here the corresponding variability and time--frequency analyses for the large-torus simulations (INSANE\_20, SANE\_20, INSANE\_0.5, and SANE\_0.5), together with the MAD\_0.5 model.

\begin{figure*}
\centering
\includegraphics[width=\textwidth]{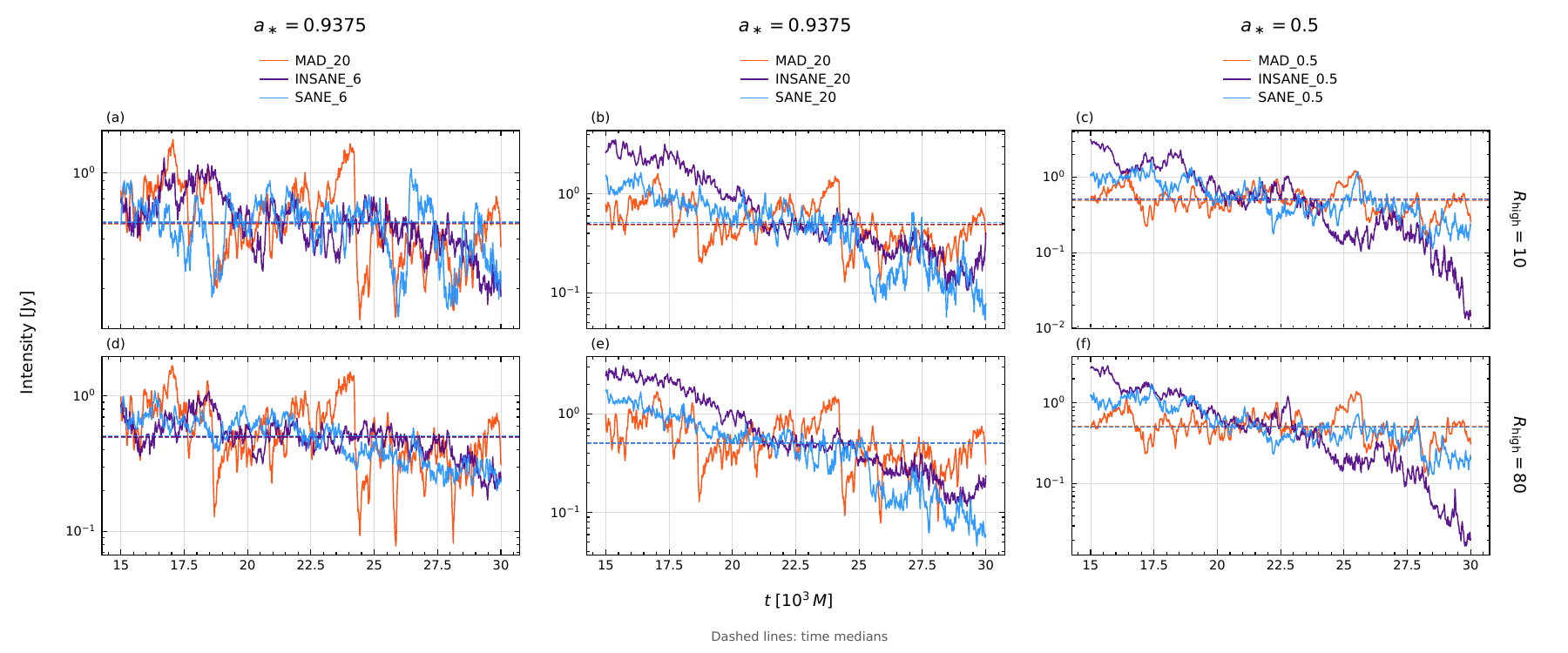}
\caption{Light curves of  the 230\,GHz total intensity $I$. MAD, INSANE, and SANE models are marked as red, purple, and blue.  The columns show, from left to right, the small-torus models with $a=0.9375$, the large-torus models with $a=0.9375$, and the models with $a=0.5$. The medians of the light curves are plotted as the horizontal dashed lines. The upper panels use the heating prescription of $R_{\rm low}=1$ and $R_{\rm high}=10$, and the lower panels use $R_{\rm low}=1$ and $R_{\rm high}=80$.}
\label{fig:lc_I}
\end{figure*}

The intensity light curves of all models are shown in Fig.~\ref{fig:lc_I}. All of them are normalized to the same median flux, but the large-torus INSANE and SANE light curves decline throughout the analysed interval. Between the first and the last quarter of the interval, the mean intensity falls by a factor of $\approx10$--$15$ for INSANE\_20 and INSANE\_0.5, by a factor of $\approx6$--$9$ for SANE\_20, and by a factor of $\approx3$ for SANE\_0.5, whereas it changes by only a factor of $\approx1.5$ for MAD\_0.5. The fractional rms variability of the large-torus INSANE light curves, 88--95 per cent, is therefore dominated by this decline and not by fluctuations about a steady state.

\begin{figure*}
\vspace*{-4.0em}
\centering
\includegraphics[width=1.9\columnwidth]{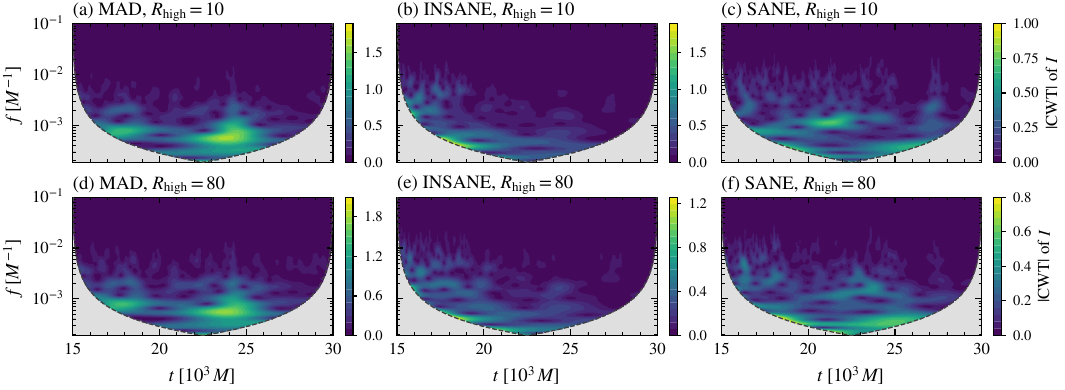}
\includegraphics[width=1.9\columnwidth]{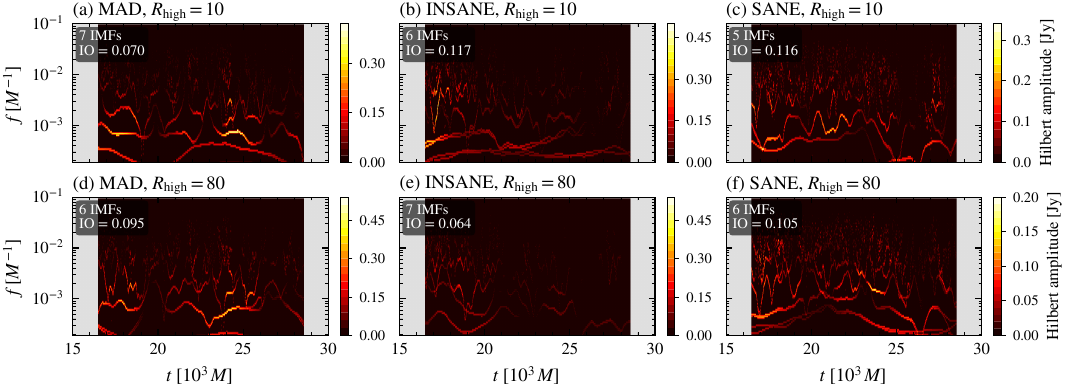}
\includegraphics[width=1.9\columnwidth]{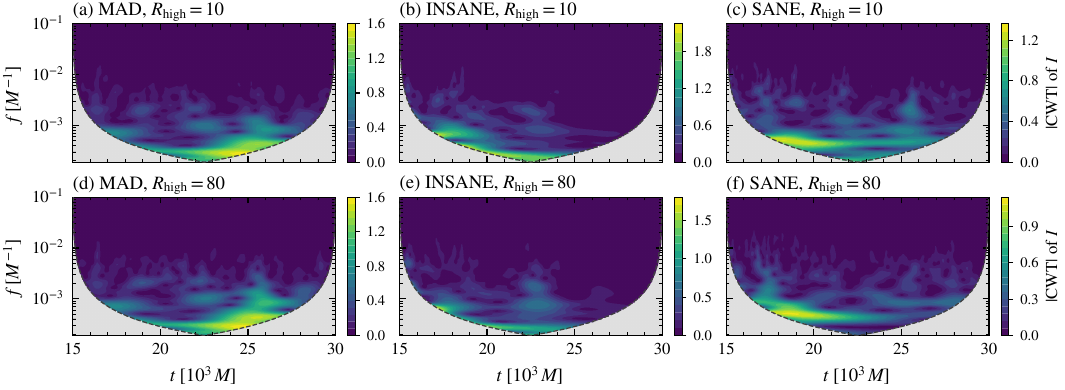}
\includegraphics[width=1.9\columnwidth]{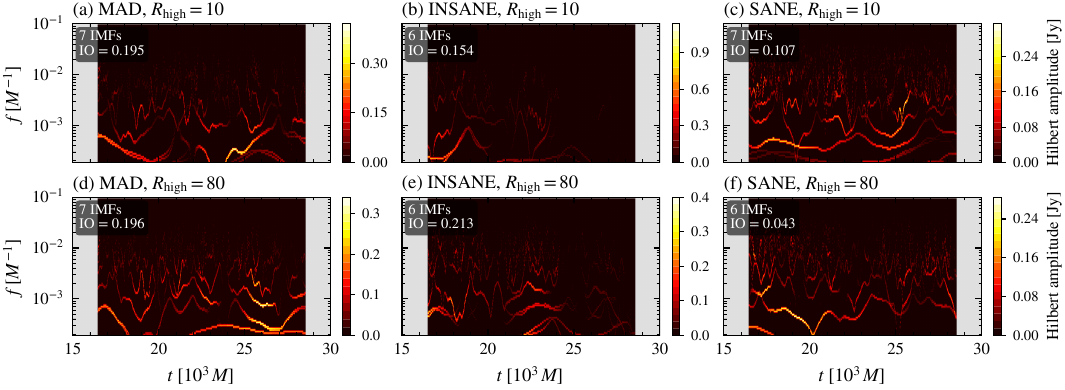}
\caption{As Fig.~\ref{fig:I_0.9375_co}, but for the large-torus models. The upper two sets of panels show the CWT scaleograms and Hilbert--Huang spectra of the $a=0.9375$ models (MAD\_20, INSANE\_20 and SANE\_20; the MAD panels are the same as in Fig.~\ref{fig:I_0.9375_co}), and the lower two sets show those of the $a=0.5$ models (MAD\_0.5, INSANE\_0.5 and SANE\_0.5). Each panel has its own colour scale.}
\label{fig:I_0.9}
\end{figure*}

The corresponding CWT and HHT analyses are shown in Fig.~\ref{fig:I_0.9}. The secular decline affects them in two ways. First, it concentrates the amplitude at the lowest frequencies. The scaleograms of INSANE\_20 and SANE\_20 are strongest along the low-frequency boundary of the cone of influence, and their time-averaged wavelet spectra are still rising at the lowest accessible frequency. In the HHT, the trend is carried by the residue together with the IMFs slower than the lowest frequency accessible to the CWT ($f<1.9\times10^{-4}\,M^{-1}$), which account for $\gtrsim85$ per cent of the variance of the large-torus INSANE light curves and for 70--91 per cent of that of the large-torus SANE light curves. Secondly, the wavelet and Hilbert amplitudes are measured in Jy and therefore scale with the flux level. In the INSANE\_20 and INSANE\_0.5 panels almost all of the structure lies at $t\lesssim21\,000\,M$, where the flux is highest, with maxima at $f\approx(3$--$6)\times10^{-4}\,M^{-1}$ and $t\approx17\,000$--$20\,000\,M$, and the panels are nearly empty at later times. This asymmetry follows the decline of the flux and does not by itself indicate a change in the character of the variability.

The MAD\_0.5 model evolves the least and behaves like the MAD model of the main text. Its wavelet amplitude is dominated by one episode at $t\approx24\,000$--$26\,000\,M$ and $f\approx(3$--$4)\times10^{-4}\,M^{-1}$ (time-scales of $\approx2300$--$3600\,M$), which corresponds to a flare of the light curve to $\approx1.2$--$1.4$\,Jy at $t\approx25\,500\,M$ followed by a rapid decline. The Hilbert spectra show their brightest ridge at the same time, and the result is insensitive to $R_{\rm high}$. The characteristic time-scale of this episode is about twice that of the corresponding episodes in the $a=0.9375$ MAD model.

\bsp	
\label{lastpage}
\end{document}